\documentclass[aps,pre,superscriptaddress]{revtex4}

\usepackage{amsfonts}
\usepackage{amsmath}
\usepackage{multirow}
\usepackage{longtable}
\usepackage{graphicx}
\usepackage{color}

\newcommand{\be}{\begin{equation}}
\newcommand{\ee}{\end{equation}}
\newcommand{\bea}{\begin{eqnarray}}
\newcommand{\eea}{\end{eqnarray}}

\begin{document}

\title{The fragility of decentralised trustless socio-technical systems}

\author{Manlio De Domenico}
\email[]{mdedomenico@fbk.eu}
\affiliation{Fondazione Bruno Kessler, 38123 Povo, Italy}

\author{Andrea Baronchelli}
\email[]{Andrea.Baronchelli.1@city.ac.uk}
\affiliation{Department of Mathematics - City, University of London. Northampton Square, London EC1V 0HB, UK}
\affiliation{UCL Centre for Blockchain Technologies, University College London, London, UK.}

\date{\today}

\begin{abstract}
The blockchain technology promises to transform finance, money and even governments. 
However, analyses of blockchain applicability and robustness typically focus on isolated systems whose actors contribute mainly by running the consensus algorithm. Here, we highlight the importance of considering trustless platforms within the broader ecosystem that includes social and communication networks. As an example, we analyse the flash-crash observed on 21st June 2017 in the Ethereum platform and show that a major phenomenon of social coordination led to a catastrophic cascade of events across several interconnected systems. We propose the concept of ``emergent centralisation'' to describe situations where a single system becomes critically important for the functioning of the whole ecosystem, and argue that such situations are likely to become more and more frequent in interconnected socio-technical systems. We anticipate that the systemic approach we propose will have implications for future assessments of trustless systems and call for the attention of policy-makers on the fragility of our interconnected and rapidly changing world.

\end{abstract}

\maketitle

\section*{Introduction}
The Internet and World Wide Web have been saluted as a major decentralising forces for at least two decades. They would give power back to the people, enabling everyone to make their voice heard and take active part in decisions. This trend continues today with major political parties in the Western World advocating the Web will enable reforming democracy ``from the bottom up''. However, the same infrastructure has allowed a few high-tech giants (such as Google, Facebook, Amazon, etc) to accumulate, and centralise, an unprecedented mix of economic, social and ultimately political power.

Recently, the blockchain technology has infused new hopes of decentralisation, promising to transform finance, money and even governments \cite{swan2015blockchain,chapron2017environment}. Its central innovation is distributed trustless consensus, which allows to determine the validity of transactions without the need of any intermediaries in a transparent and secure way \cite{nakamoto2008bitcoin}. The tremendous interest attracted by Bitcoin and other cryptocurrencies in the last few years show the potential impact of such a technology. However, the current debate on centralisation versus decentralisation often overlooks the complexity of the socio-technical ecology we inhabit, and the fact that it is ultimately dominated by human behaviour. We argue that considering this broader perspective is crucial, and call for the attention of policy-makers on the fragility of our interconnected world.

\section*{Cryptocurrencies and the Decentralisation of Money}

\begin{figure}
\begin{center}
\includegraphics[width=0.6\textwidth]{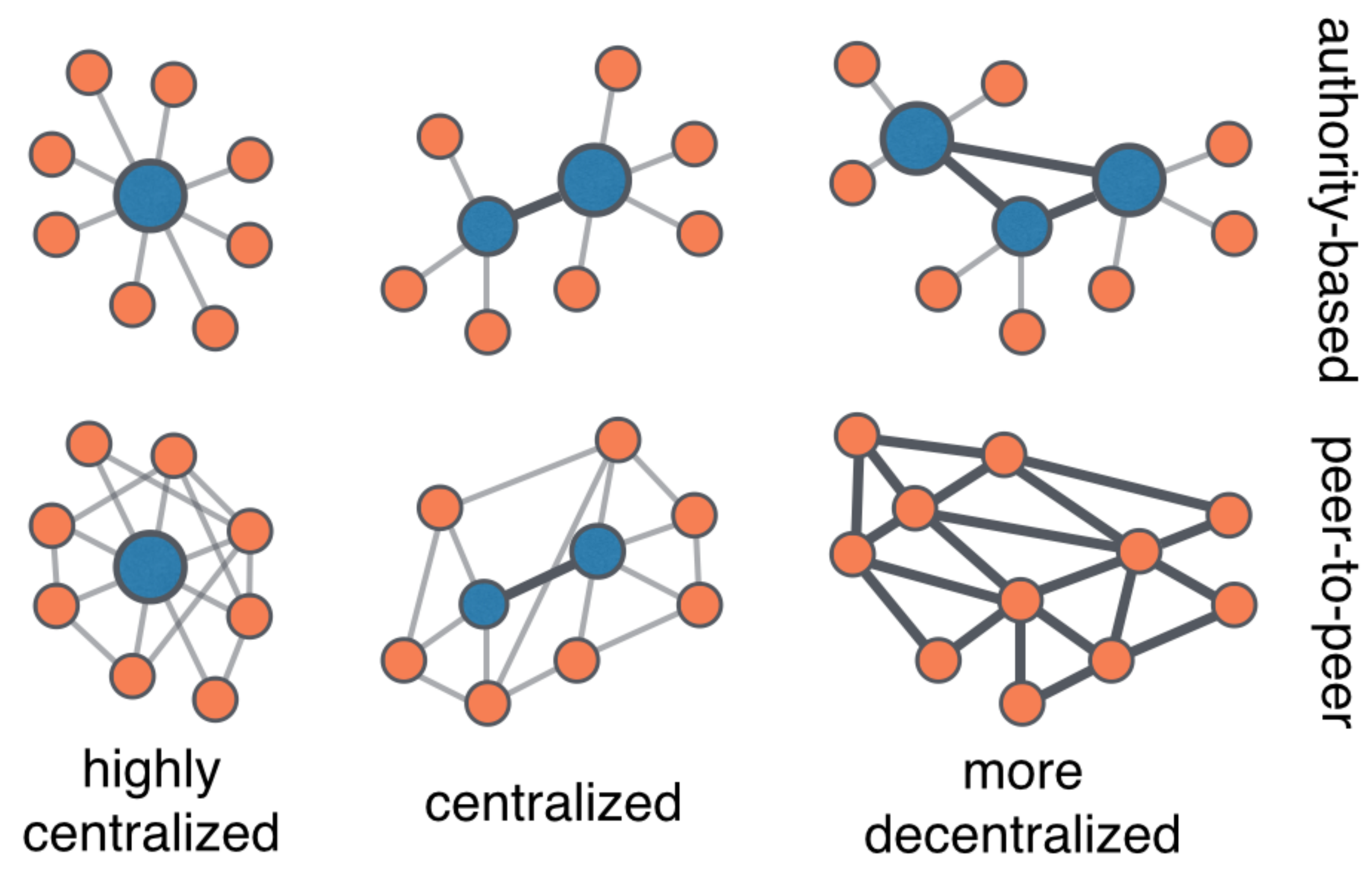} 
\end{center}
\caption{\label{fig1}\textbf{From centralized authorities to decentralized peer-to-peer networks.} In a system based on authorities, clients connect to a central node to perform an action, like sending money to other users. This type of system can be highly centralized, with just one authority in charge for providing the service, or more decentralized, with a few interconnected authorities taking care of their clients. In a peer-to-peer network, users might be still connected to authorities, although they are also directly connected each other, like in sharing systems. }
\end{figure}

Traditional digital payment systems require users, individuals or companies, to use intermediaries such as banks to circulate money. If Alice wants to send \$100 to Bob through the Internet, she has to move that amount to his banking account and tell her bank about Bob's account details. Alice's bank moves the money to Bob's bank where eventually he can withdraw the \$100. This procedure is made easy by smartphones and online applications, but relies on the central role of banks, which get paid for their services. In fact, this is an example of how a centralised system works (Figure 1).

The technology offered by Bitcoin \cite{nakamoto2008bitcoin} and other digital currencies - the blockchain - allows Alice to transfer her money directly to Bob, without centralised intermediaries. Instead, the transaction is sent to the whole network of nodes which builds, for instance, the Bitcoin system: those nodes, called miners, use computational power to solve complex mathematical problems in order to verify if the transaction is valid and consequently vote in favour or against it. Once a consensus about validity is reached, the transaction is accepted and Bob can spend those \$100 after a few minutes. The system generates some new bitcoins to reward the miners for their work.

Thus, networks based on digital currency are \textit{distributed}, because transactions are processed by multiple nodes, and \textit{decentralised}, not relying on any authority or middleman. In practice, they reduce the need to trust users or third parties during financial operations. As other decentralised infrastructures and protocols \cite{de2017modeling}, these networks are very robust to perturbations, such as node's failure and targeted attacks that, conversely, might significantly damage other interconnected centralised systems like the financial system or airline traffic \cite{albert2000error}. One of the most remarkable feature of digital currency is pseudo-anonymity, sometimes exploited for illicit purposes through the Dark Web. 

Over time, thousands of new projects similar in spirit to, or heavily based on, the Bitcoin system appeared and created a growing ecology of cryptocurrencies \cite{elbahrawy2017evolutionary}. At the moment of writing, more than 1,600 cryptocurrencies can be traded, at least in theory, totalling a market cap value of almost \$~150billions. Tens of banks -- including Accenture, Bank of England, Santander, UBS and UniCredit -- have publicly integrated or are experimenting the blockchain technology, whereas companies like JP Morgan have recently started partnerships with blockchain platforms like ZCash to improve the privacy of their customers.

\begin{figure}
\begin{center}
\includegraphics[width=0.9\textwidth]{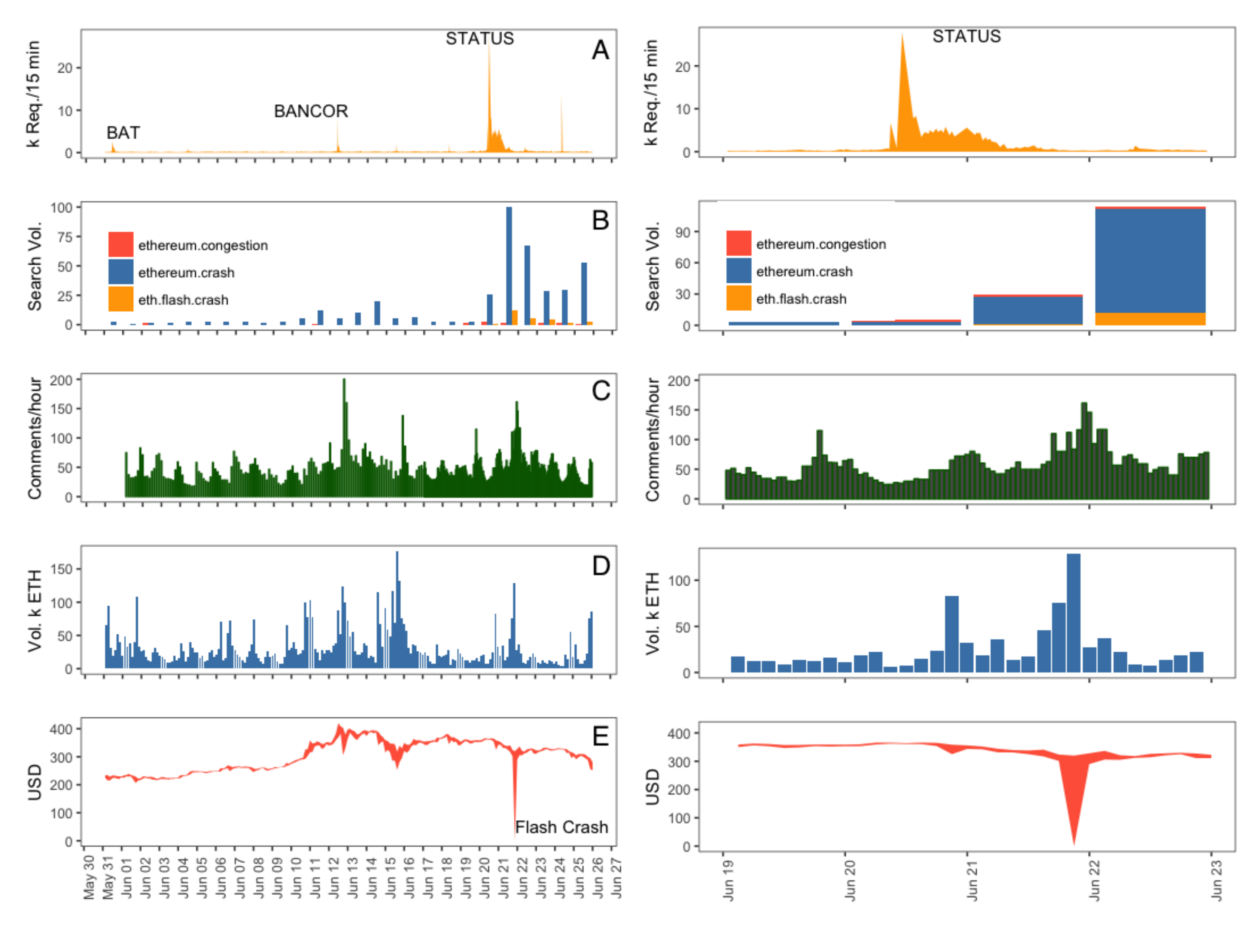} 
\end{center}
    \caption{\label{fig2}\textbf{Dynamics of different interdependent systems} \emph{Left}: (A) Raw number of transaction requests (in thousands) to the Ethereum network (specifically, through myetherwallet.com) over time, with notable peaks corresponding to some well known ICOs. (B) Search volume of Google queries related to congestion and crash of the Ethereum system. (C) Number of comments per hour on Reddit about Ethereum (D) Volume of ``ethers'' (in thousands) traded on GDAX. (E) Low and high values of ethers exchanged for US dollars on GDAX (data from cryptocompare.com). \emph{Right:} details around the Flash Crash event.}
\end{figure}

\section*{Ethereum, A Platform For Smart Contracts}

The blockchain technology can be used also for other type of interactions where trust is usually required. This groundbreaking intuition has led to the technology of ``smart contracts'' provided by the non-profit distributed computing platform Ethereum. The underlying idea is that many contractual clauses can be made fully-executing and self-enforcing as the terms of the agreement between parties is translated into lines of code. In practice, smart contracts are high-level programming abstractions that are deployed to the Ethereum blockchain for execution. Once on the blockchain, the coded contract is temper-resistant, self-verifying and self-enforcing. Smart contracts also play the role of `API connectors' to the blockchain and have enabled the emergence of `blockchain enabled' serverless websites, known as decentralised applications (Dapps).

The Ethereum technology is quickly gaining consensus among developers as the reference platform for the implementation of a socio-technical ecosystem where technological applications and services are no more based on trust among parties. Hundreds of new technological companies nowadays propose services based on smart contracts, together with innovative solutions for user privacy and data management with high potential impact on society and technology. These companies often attract investors through crowdfunding operations known as initial coin offering (ICO). During an ICO, tokens -- instead of shares, as in IPOs -- are sold before the corresponding Dapp is released. ICOs are still not regulated like IPOs and token-holders, as opposed to share-holders, have little or no decisional power at all in the company.

\begin{figure}
\begin{center}
\includegraphics[width=0.8\textwidth]{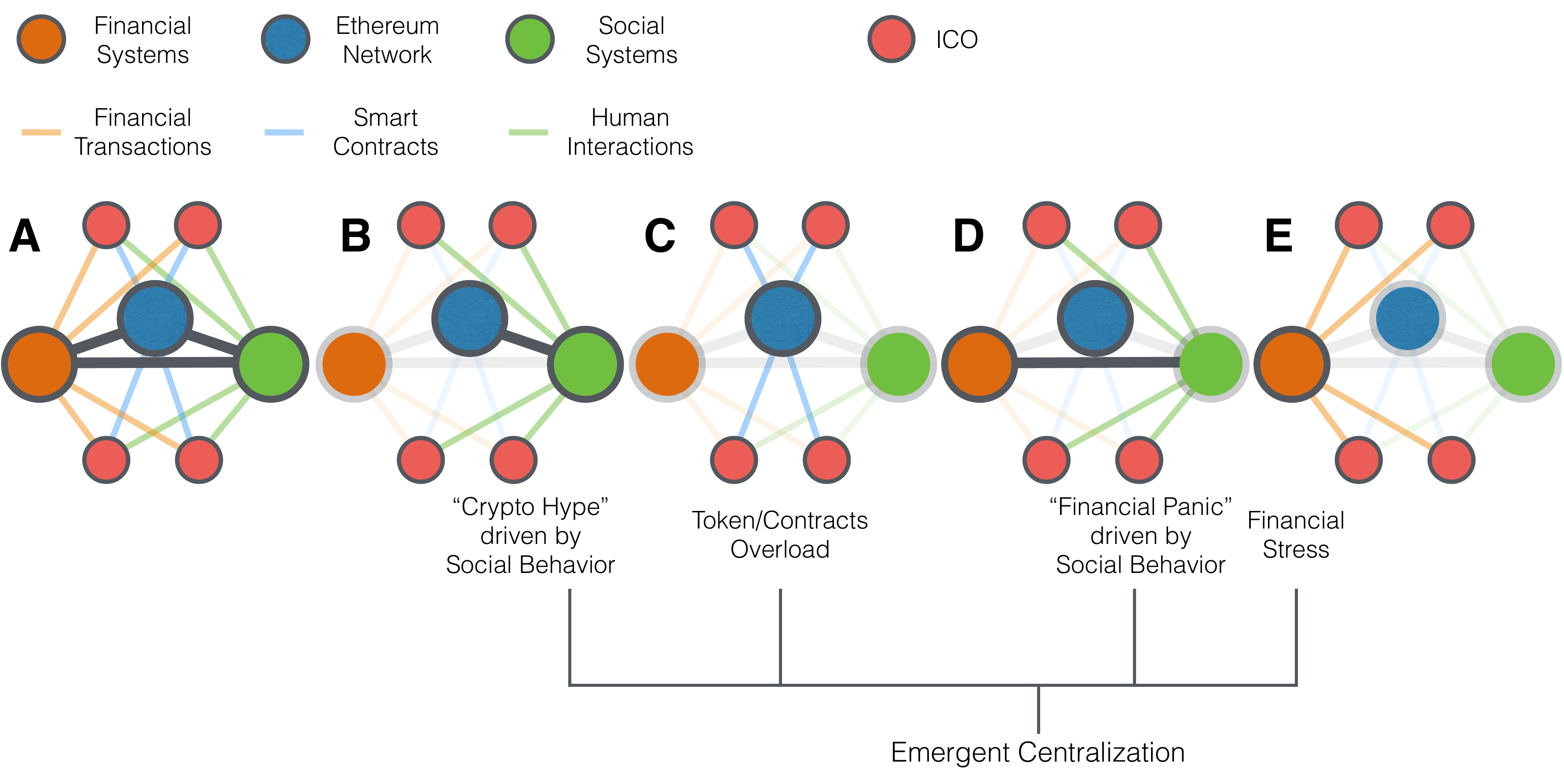} 
\end{center}
\caption{\label{fig3}\textbf{Emergent centralization.} (A) The ecosystem of socio-technical systems involving the Ethereum network, online social networks and financial systems (e.g., exchanges). (B) Social networks drive the hype about ICOs. (C) The Ethereum network is overloaded by investors and the system slows down. (D) Fear for a (supposed) crash generates a financial panic leading to a (E) stress of financial systems where, eventually, failures are expected.}
\end{figure}

\section*{Blockchains Do Not Live in Isolation}

The tumultuous innovation rate in the world of blockchain inevitably draws most of the attention on the latest novelty. However, we believe it is crucial to broaden the perspective and consider the single platform as one element in the ecosystem of socio-technical systems we live in. The recent ``flash crash'' experienced by GDAX --  one of the biggest cryptocurrency exchanges -- is a good example to illustrate this. On June 21, 2017 the value of 1 ether -- the digital currency of Ethereum -- crashed from \$320 to \$0.1 for a few seconds. Although investigations are undergoing, the mainstream reporting explained that the crash was caused by a sell order at market price worth several million dollars, causing the exchange to accept bids at any price to fill the request, which in turn dropped the value of 1 ether to \$224. Such a rapid change has triggered a cascade of hundreds of automatic margin calls and stop loss orders set by traders, which eventually have driven the price to \$0.1 \cite{flash_crash}. In summary, the event originated from a single major player - it was a centralised event. As simple as it seems, this is just an approximated description of what happened and placing Ethereum in a broader context is crucial to clarify the nature of the crash. 

The Ethereum crash was a major phenomenon of social coordination leading to a catastrophic cascade of events across interconnected systems.
Ethereum is the central node of the ecosystem of platforms fuelled by ICOs: every time an ICO is taking place, thousands of transactions calls are made by investors to buy tokens. Such calls have to be executed and verified by miners, stressing the Ethereum network well above its normal workload. By inspecting the dynamics of raw transaction requests to the system (Figure 2A) it is evident that each ICO generates at least 10 times more requests than normal, with the ICO of STATUS (on June 20, 2017) generating an unexpected traffic which has paralysed the system for hours. Buyers of tokens and standard users experienced a huge delay, with most of them starting to worry about an Ethereum crash, as shown by the sudden increase in the amount of Google queries about system's crash and congestion (Figure 2B). Compared to the few minutes required to validate requests, the network was not responsive for several hours, and the news started to propagate across social networks like Reddit (Figure 2C). The result has been the emergence of collective panic, spread also through other platforms, leading big holders to sell their ethers before they would lose their value because of the (supposed) system's crash (Figure 2D). This action eventually led to the flash crash at the end of June 21, 2017 (Figure 2E).

\section*{A Network of Networks}

Far from being an isolated case, the June flash crash is just one example of many similar events. Among these, the recent case of CryptoKitties highlights the central role of the human factor in the dynamics of the whole system \cite{consensus}. CryptoKitties is a blockchain-based game that allows players to adopt, raise, and trade virtual cats, and it represents one of the first successful attempts to use blockchain technology for recreational purposes. Virtual cats have been sold for more than \$100,000 and the game's popularity in December 2017 caused the Ethereum network to slow down significantly \cite{bbc_crypto_gatti}.

Such dynamics appear natural by adopting an integrated complex systems perspective. In fact, Ethereum is playing the role of central node in a system of interdependent systems, including smart-contract-based platforms, trading and - crucially - social networks. The result is a network of networks \cite{gao2011networks} characterised by distinct structural and dynamical properties \cite{de2016physics} which typically make those systems even more fragile than simpler networks \cite{buldyrev2010catastrophic} and more exposed to systemic risks. In the case of Ethereum, the traffic stress caused from other platforms generates delays or cascading failures \cite{yang2017small} that might worry people, who in turn are part of those social and trading systems where information is quickly propagated or distorted in echo chambers \cite{del2016spreading}, causing extreme reactions (e.g., placing big sell orders).

\section*{Conclusion}

The Ethereum flash crash illustrates how the decentralised nature of one system (the Ethereum blockchain, in the example) does not prevent the emergence of centralisation at a higher level. As an ecosystem grows around a blockchain based system, the latter acquires a central role in connecting the social, financial and technological elements at play.  Feedback loops and non-linearities, as well as targeted attacks, can then not only put at risk the decentralisation of the original system (e.g., in the simplest book example, polls of miners could coordinate malicious actions on a social network) but also affect the stability of the whole socio-technical ecology (Fig.~\ref{fig3}). We anticipate that such phenomena of  ``emergent centralisation'' will increase rapidly in the near future.

The fragility of emerging systems of systems is still to be understood, opening uncountable and exciting research opportunities with high societal impact.  Minimising systemic risks is crucial and requires accounting for (i) the rapidity of information propagation in online social networks, (ii) the almost instantaneous response of automated systems and sub-second networks \cite{johnson2017slow} and (iii) the interdependent nature of these network of networks. New science-driven policies are needed to guarantee that existent and forthcoming ecosystems become effective, efficient and resilient. We think that taking a systemic approach will be crucial for future assessments of trustless systems, and call for the attention of policy-makers on the fragility of our interconnected and rapidly changing world.

\section*{Acknowledgements}
The authors thank the owner of the Web service ``myetherwallet.com'' for providing the raw number of transaction requests to the Ethereum network.



\end{document}